\documentclass{article}
\usepackage{amssymb,amsmath}

\begin{document}
\title{text}

\title {Reappraisal of the causal interpretation of quantum mechanics and of the quantum potential concept }

\author {R. H. Parmenter \\ Department of Physics, University of Arizona, Tucson, AZ 85721 \and Andrew L. DiRienzo \\ 118 Weaver Road, Elizaville, NY 12523 andrew@dirienzo.org }

\maketitle

\begin{abstract}

The causal interpretation of quantum mechanics, as originally
stated by deBroglie and  Bohm, had several attractive features.
Among these is the possibility that it could address some of the
most fundamental questions on quantum phenomena. However,
subsequent theoretical conjectures, which have now been included
in the orthodox view of the deBroglie Bohm theory, are unphysical
and have done much to undermine the original theory's appeal. We,
therefore, return to the original theory as our starting point and
address one of its perplexing areas: the quantum potential. By
avoiding the unphysical conjectures we are led to an understanding
of the quantum potential which is distinctly different from that
of the orthodox deBroglie Bohm view.

PACS:  03:65


\end{abstract}

\section{Introduction}

The essential postulates of the deBroglie Bohm causal theory [1,2]
can be summarized as:

\begin{itemize}
  \item An individual physical system comprises a wave together with a
point particle. The wave and the particle are distinct parts of
the same system. The wave and the particle are never separated.
  \item The wave $\psi$ is a solution to the Schr\"{o}dinger equation.
  \item Particle motion is obtained as the solution to a modified
Hamilton - Jacobi equation. The modification to the classical
Hamilton - Jacobi equation is the addition of a term Q, the
quantum potential. The functional form of Q is determined
mathematically by $\psi$.
\end{itemize}

The causal theory has many attractive features. For example:
\begin{itemize}
  \item It is deterministic.
  \item It allows for the possibility of a complete description
of a physical system, i.e., it addresses one of the fundamental
complaints Einstein had about the Copenhagen interpretation [Ref.
2, pp. 11-15].
  \item Particle dynamics is written as the sum of classical
and quantum terms [Ref. 1, p. 29].
  \item There is a natural relationship between the
classical and quantum regimes [Ref. 2, pp. 270-274].
\end{itemize}

Together these allow for the possibility that the Causal Theory
may be able to address some of the deepest problems associated
with the Copenhagen interpretation. The theory may also lead to
new physics.

There are, however, weaknesses in the original theory. One of the
most obvious of these relates to the quantum potential Q: What is
its source? Typically in physics a force, and its associated
potential, have a source. However, nowhere in the literature is
this fundamental question addressed in a physically reasonable
way.

To compound the problem, the adherents of the theory have gone
further and made conjectures leading to results which are in
direct contradiction with experiment.

These conjectures, which we will discuss later, have now become
part of the orthodox view of the deBroglie Bohm causal
interpretation. Unfortunately, they have done much to undermine
the appealing aspects of the original theory.

Our approach is as follows:

\begin{enumerate}
  \item Return to the original theory as stated by \mbox{deBroglie} and Bohm.
  \item Reject all unphysical conjectures.
  \item Let experiment guide us as much as possible.
  \item For those things about which Copenhagen makes a
prediction, our theory must be in agreement.
  \item Start with a many-body approach because it is
closer to physical reality than a one-body approach.
  \item Address two fundamental problems: · What is the physical reason for the
forces associated with the quantum potential? ·   What is the
source of these quantum forces?
\end{enumerate}

Our goal is to then go on in future work to see if the causal
theory can be used to address some of the compelling questions of
the our day.

\section{Basics Of The Causal Theory}

We start by considering a physical system consisting of  n
interacting particles. The time dependent Schr\"{o}dinger equation
for this system is,

\begin{eqnarray}\imath\hbar\left(\frac{\partial\psi}{\partial
t}\right)=\left[\sum_{i=1}^{n}\left(\frac{-\hbar^{2}}{2m_{i}}\right)\nabla_{i}^{2}
+ V \right]\psi\end{eqnarray} where
\[\psi = \psi (\textbf{x}_{1}, \textbf{x}_{2}, \textbf{x}_{3}, . . . \textbf{x}_{n},
t),\]\
\[ V = V(\textbf{x}_{1}, \textbf{x}_{2}, \textbf{x}_{3}, . . . \textbf{x}_{n},
t),\]\
\[\nabla_{i} = \left( \frac{\partial}{\partial x_{i}},\frac{\partial}{\partial y_{i}},\frac{\partial}{\partial z_{i}}\right),\] \\ and $(\textbf{x}_{1},\textbf{x}_{2}, \textbf{x}_{3}, . . .
\textbf{x}_{n})$ provides a set of rectangular Cartesian
coordinates of the n particles. Also V is the classical potential
energy which includes the interparticle and external potentials
from the traditional forces (gravity, E\&M, strong and weak
interactions.)

Assume that the energy spectrum is discrete, so that the
wavefunction  $\psi$  is localized in configuration space. Without
loss of generality, the wavefunction can be written as
\begin{eqnarray} \psi = R\exp \imath (S/\hbar) \end{eqnarray}
where  R  and   S  are real and
\[R = R (\textbf{x}_{1}, \textbf{x}_{2}, \textbf{x}_{3}, . . . \textbf{x}_{n}, t)\]
\[S = S (\textbf{x}_{1}, \textbf{x}_{2}, \textbf{x}_{3}, . . . \textbf{x}_{n}, t)\]
Equation (1) is equivalent to the pair of equations
\begin{eqnarray}\frac{\partial S}{\partial t} + \sum ^{n} _{i = 1} (\nabla_{i}
S)^{2} (2m_{i})^{-1} + Q + V = 0 \end{eqnarray}

\begin{eqnarray}\frac{\partial R^{2}}{\partial t} + \sum ^{n} _{i =
1}\nabla_{i}\cdot \left(\frac{R^{2}\nabla_{i} S}{m_{i}}\right)= 0
\end{eqnarray}
where \[Q = Q (\textbf{x}_{1}, \textbf{x}_{2}, \textbf{x}_{3}, . .
. \textbf{x}_{n}, t)\]
\begin{eqnarray} Q = - \sum ^{n} _{i =
1}\left(\frac{\hbar^{2}}{2m_{i}R}\right)\nabla_{i}^{2}R
\end{eqnarray} \\ is defined as the quantum potential [Ref. 2, p. 279]. If it were
not for the presence of the quantum potential Q,  Eq. (3) would be
the classical Hamilton - Jacobi equation.

Just as the causal form of the Hamilton - Jacobi equation contains
the additional term Q, the causal form of Newton's second law
contains an additional term involving Q [Ref. 2, p. 279-280],
\begin{eqnarray}\frac{d\bold P_{i}}{dt}= - \bold\nabla_{i}V - \nabla_{i}Q
\end{eqnarray}
here $ \textbf{P}_{i}$ is the momentum of the i-th particle, $ -
\nabla V_{i} $ is the sum of all the conventional forces on the
i-th particle, and $- \nabla_{i}Q $ is interpreted as the quantum
force on the i-th particle. The total momentum of our n-particle
system is given by
\begin{eqnarray}\ \bold P = \sum_{i = 1}^{n} \bold P_{i}\end{eqnarray}
with dynamical equation [Ref. 2, p. 285]
\begin{eqnarray} \frac{d\bold P}{dt}= - \sum^{n}_{i=1}\nabla_{i}V - \sum^{n}_{i =
i}\nabla_{i}Q
\end{eqnarray}\\

\section{Studies of the quantum potential}

The purpose of this paper is to attempt to answer the following
questions. What is the physical reason for the forces associated
with the quantum potential? Furthermore, what is the source of
these forces? We suggest that Q is the result of a nonholonomic
nonlocal constraint on the system, the requirement that the
wavefunction $ \psi $    not exhibit deterministic chaos (extreme
sensitivity to initial conditions.) The forces associated with Q
are forces of constraint. And we furthermore suggest that the
source of the quantum force on a given particle is the other
particles of our n particle system.

Why are we interested in preventing deterministic chaos? In
general, quantum mechanical systems localized in configuration
space are described by Hermitian Hamiltonians. With such a
Hamiltonian, the wavefunction is never subjected to extreme
sensitivity to initial conditions. However, removing Q from the
causal Hamilton-Jacobi equation (Eq. 3) opens the possibility that
the corresponding wavefunction be subjected to extreme sensitivity
to initial conditions. All this is shown in Appendix A. In
Appendix B ways to detect the presence of deterministic chaos are
discussed.

While deterministic chaos of the wavefunction should not occur in
a localized quantum system, it is quite possible that such chaos
can be associated with the particle trajectories of the system, as
calculated by deBroglie Bohm theory [4].  Furthermore, in a
macroscopic quantum system the wavefunction may exhibit
deterministic chaos, as is discussed in Appendix C.

We now turn to our second question: What is the source of the
quantum forces? On this issue the leading proponents of the causal
theory make an assumption (conjecture) which we find truly
mysterious. Consider the following quotes:

\begin{enumerate}
  \item On page 170 of  Ref.3, Bohm states that the field $\psi$
``exerts a force on the particle in a way that is analogous to,
but not identical with, the way in which an electromagnetic field
exerts a force on a charge, and a meson exerts a force on a
nucleon\footnote{It should be noted that Bohm and Hiley [Ref. 1,
pp. 29, 30, 37, 38, 40] imply, in contrast, that in our Eq. (6) ,
$- \nabla_{i}Q $ is a force being exerted on the i-th particle,
but that it is not directly exerted by the wavefunction $ \psi $.
Rather, the wavefunction controls the action and reaction of the
particle with something like the fluctuations of the vacuum,
causing ``acceleration of the particle in its self-movement.''}.''
  \item On page 91 of Ref. 2, Holland states ``it should be emphasized that
while the quantum field \mbox{[ $\psi$   ]}  does not push on the
particle as we might expect a classical wave to, it does
nevertheless guide the particle by exerting a direct force on it
via Q.'' \\
\\
We infer from these statements that Bohm and Holland make the
assumption that the wavefunction, not only is an independent
entity distinct from the particle itself, but also is the "source"
of the quantum force!  Consequently, it is no surprise when
Holland goes on to say :

  \item Ref. 2, p. 79: ``the particle simply responds to local values of
the field in the vicinity  (via Q) - there is no reciprocal action
of the particle on the wave.''
  \item  Ref. 2, p. 286:  ``A classically closed system may not be closed
quantum mechanically.''
\end{enumerate}
\[ \frac{d\textbf{P}}{dt}=  - \sum^{n}_{i = 1}\nabla_{i}Q \neq 0
\]

The assumption that the source of the quantum force is the
wavefunction itself has led the orthodox proponents of the theory
to unphysical results. For example, the third quote above suggests
that Newton's third law (to every action there is an equal and
opposite reaction) is violated. Furthermore, the fourth quote
allows for the possibility that an isolated system may
self-accelerate in the absence of any known force. Not only are
these results in disagreement with predictions made using the
Copenhagen interpretation of quantum mechanics, but they are also
in disagreement with all known experiments.

These unphysical results are a consequence of combining the
assumption that ``the wavefunction itself exerts a force on the
particle" with the fact that the Schr\"{o}dinger equation is
homogeneous. Rather than questioning the assumption on the
wavefunction,  it is apparently the orthodox deBroglie Bohm
approach to accept it and instead speculate on the possibility of
some inhomogeneous version of the Schr\"{o}dinger
equation\footnote{ Bohm [Ref. 3, p. 179] suggests that at a
distance less than $ 10^{-13}$ cm, Schr\"{o}dinger's equation
becomes inhomogeneous allowing a particle to react back on the
field, in analogy with the inhomogeneous Maxwell's equations
allowing a charge to act on the electromagnetic field.}. There is,
however, no experimental evidence to support these speculations.

We take a different approach. To obtain a physically reasonable
causal theory we add two assumptions to the original deBroglie
Bohm postulates:

\begin{enumerate}
  \item The wavefunction exerts no forces on the particles of
  the system. Although it is apparent from Eq. (1) through (5) that the wavefunction determines the functional form of the quantum potential, there is no indication that $ \psi $  itself exerts forces on the particles of the system.
  \item For an isolated system, the total momentum $\textbf{P}$ is a constant of the motion; i.e., it does not change with time. This is consistent with experiment. It is also consistent with the Copenhagen interpretation; because for an isolated system, the wavefunction is an eigenfunction of the total momentum operator.
\end{enumerate}

Using these ideas, and returning to our many-body equations, we
are led to a very different result for the source of the quantum
force than that in the orthodox deBroglie Bohm view.

Consider the situation in which our n-particle system is an
isolated system.  Since the traditional forces obey Newton's third
law we have
\[  - \sum^{n}_{i = 1}\nabla_{i}V = 0 \]
For an isolated system we have
\[ \frac{d\textbf{P}}{dt} = 0 \]
so that
\[  - \sum^{n}_{i = 1}\nabla_{i}Q = - \sum^{n}_{i = 1}\textbf{F}_{i}= 0 \]
That is, the sum of all the quantum forces on the n-particles is
zero. From this we have
\begin{eqnarray} \textbf{F}_{i} = \sum_{j \neq i}^{n} ( -
\textbf{F}_{j}) \end{eqnarray}
$ \textbf{F}_{i}$   is the quantum
force on the i-th particle. Eq. (9) states that the quantum force
on the  i-th particle of our isolated system is equal and opposite
to the sum of the quantum forces on all the other particles. For
\mbox{n = 2}, this gives $\textbf{F}_{1} = - \textbf{F}_{2} $,
i.e., the quantum force on particle 1 is equal and opposite to the
quantum force on particle 2. This strongly suggests that the
source of the quantum force on one particle is the other particle.
(Which, conveniently, is also consistent with Newton's Third Law.)

Eq. (9) implies that the net quantum force on the i-th particle is
the result of all the other particles of the system exerting
forces on the particle through the intermediary of the quantum
potential. If the system is acted on by external forces, the
simplest thing to do conceptually is to enlarge the system to
include the bodies creating the external forces, so that one has
an enlarged isolated system. We see that the source of the quantum
force on any particle is nothing more that all the other particles
of the system. To us, this seems to be a much more satisfying
explanation than is that of Holland or of Bohm and Hiley [Ref. 1,
pp. 30, 37, 38, 40]. The nonlocality of the quantum potential
discussed by Holland [Ref. 2, p. 282] is the cause of the nonlocal
interactions between the particles of the system.

Consider the situation of an isolated hydrogen atom in the ground
state. The total force on the atom vanishes. At the same time
there are equal and opposite Coulomb forces of attraction on the
electron and on the proton. Since the internal wavefunction of the
system is real, the causal theory says that the electron is
motionless relative to the proton. This results from the fact that
there are equal and opposite quantum forces of repulsion between
the two particles, such that there is no net force on either
particle; and their relative position stays fixed.

With this understanding of the physical reason for, and the source
of, the quantum forces, the authors feel that the concept of the
quantum potential is less strange and more reasonable than would
appear at first sight.

\section{Acknowledgments}
The one of the authors (ALD) would like to thank Harold Brooks and
Richard Shapiro for several helpful discussions.

\section{Appendix A}

When we remove Q  from the Hamilton - Jacobi equation   [Eq. (3)],
we are adding to the Schr\"{o}dinger equation the term [Ref. 2, p.
56]

\begin{eqnarray} \left(\frac{\hbar^{2}}{2m}\right) \exp \imath (S/\hbar)
\nabla^{2}R & = & \left(\frac{\hbar^{2}}{2m}\right)\mid \psi \mid
^{-1} \psi \nabla^{2}\mid \psi \mid \nonumber \\
&  &  \nonumber \\
& = & - Q \psi \end{eqnarray}\\so that the effective Hamiltonian
becomes

 \begin{eqnarray}H_{eff}= - \left(\frac{\hbar^{2}}{2m}\right)\nabla^{2} + V +
\left(\frac{\hbar^{2}}{2m}\right)\mid \psi \mid ^{-1}
\nabla^{2}\mid \psi \mid\end{eqnarray}

Note that   $H_{eff}  =  H_{eff} ( \psi  ) $ is a functional of
$\psi$, the state upon which it is operating.  This means that the
superposition principle no longer holds.   When  $ \varphi \neq
\psi $,

\begin{eqnarray} \lefteqn{\int (\varphi^{*}H_{eff}\psi - \psi
H_{eff}^{*}\varphi^{*})d\tau }   \\
&  &  \nonumber\\
&  = & \left(\frac{\hbar^{2}}{2m}\right)\int\varphi^{*}\psi \,
\Big[ \mid \psi \mid ^{-1} \nabla^{2}\mid \psi \mid  - \mid
\varphi \mid
^{-1} \nabla^{2}\mid \varphi \mid \, \Big] d\tau \nonumber  \\
&  &  \nonumber \\
& \neq &  0   \nonumber   \end{eqnarray}\\so that  $H_{eff}$   is
not Hermitian, $ H_{eff}\neq H_{eff}^{\dag} $. This means that the
time development operator $\exp [(\imath/\hbar)H_{eff}t]$   is not
unitary.  The time dependent Schr\"{o}dinger equation is a
\textit{nonunitary} flow. Since

\begin{eqnarray}\imath\hbar\left(\frac{\partial}{\partial
t}\right)\psi^{*}\psi = \psi^{*}H_{eff}\psi - \psi
H_{eff}\psi^{*}\end{eqnarray} \\we have

\begin{eqnarray} \lefteqn{\imath\left(\frac{\partial}{\partial t}\right)\int\mid \psi
\mid^{2}d\tau = } \\
&  &  \nonumber\\
&  & \left(\frac{\hbar}{2m}\right) \int \mid \psi
\mid^{2}\Big[\mid \psi \mid^{-1}\nabla^{2}\mid \psi \mid - \mid
\psi \mid^{-1}\nabla^{2}\mid \psi \mid\Big]d\tau \nonumber
\end{eqnarray}\\and the normalization of the wavefunction is time independent.
When $\varphi \neq \psi $

\begin{eqnarray} \lefteqn {\imath \left(\frac{\partial}{\partial t}\right)\int
\varphi^{*}\psi d\tau} \\
&   &  \nonumber\\
&  & = \left(\frac{\hbar}{2m}\right)\int \varphi^{*}\psi\Big[\mid
\psi \mid^{-1}\nabla^{2}\mid \psi \mid - \mid \varphi
\mid^{-1}\nabla^{2}\mid \varphi \mid \Big]d\tau \nonumber \\
&   &  \nonumber\\
&  & = 0 \nonumber \end{eqnarray}\\so that

\begin{eqnarray} \lefteqn{\left(\frac{\partial}{\partial t}\right)\int\mid \psi -
\varphi\mid^{2}d\tau } \\
&  &  \nonumber\\
&  &  = \left(\frac{\hbar}{2m}\right)\int \imath
\big[\psi^{*}\varphi - \varphi^{*}\psi\big] \big[\mid \psi
\mid^{-1}\nabla^{2}\mid \psi \mid - \mid \varphi
\mid^{-1}\nabla^{2}\mid \varphi \mid \big]d\tau \nonumber\\
&  &  \nonumber\\
&  & \neq 0 \nonumber
\end{eqnarray}\\The normalization of $ ( \psi - \varphi )$   is real and time
dependent.

Consider the case where two initial conditions for the time
dependent Schr\"{o}dinger equation differ only infinitesimally. As
time progresses the two corresponding wavefunctions can become
quite different. This indicates the possibility of deterministic
chaos of the wavefunction (extreme sensitivity to initial
conditions.) All this is a consequence of $ H_{eff}$ being a
functional of the state upon which it is acting.

If we remove the term

\[ \left(\frac{\hbar^{2}}{2m}\right)\mid \psi\mid ^{-1}\nabla^{2}\mid \psi\mid
\]\\from Eq. (11) , we are left with a Hermitian Hamiltonian, as
demonstrated by removing the corresponding terms in Eq. (12).
Removing the corresponding terms from Eqs. (13) and (14) , we see
that the normalization of $ ( \psi - \varphi)$ is time
independent. This shows that there can be no deterministic chaos
of the wavefunction associated with a Hermitian Hamiltonian.

\section{Appendix B}

In order to detect the presence of deterministic chaos, it is
necessary to transform the time dependent Schr\"{o}dinger equation
so that it looks like a set of classical equations of motion [5].
By expanding the wavefunction  $ \psi( t ) $ in terms of the
members of some orthonormal complete set  $\varphi_{i}$

\begin{eqnarray} \psi(t) =
\sum_{j}a_{j}(t)\varphi_{j}\end{eqnarray}\\the Schr\"{o}dinger
equation becomes

\begin{eqnarray}\frac{da_{j}}{dt}=
\sum_{k}M_{jk}a_{k}\end{eqnarray}\\with

\begin{eqnarray} M_{jk}\equiv (\imath\hbar)^{-1}\langle j\mid H \mid k
\rangle.
\end{eqnarray}

This is a  complex  N  dimensional flow, where N is the dimension
of the Hilbert space appropriate to the problem.  (The equivalent
real flow is 2N dimensional.)  Under ordinary circumstances the
flow is linear and there is no reason to expect deterministic
chaos. However, if the Hamiltonian is a functional of the state of
the system, then  $M_{jk}$ will be a function of the coefficients
$ a_{i}$ and the flow is \textit{nonlinear}. Thus, if  $2N > 3 $,
the Poincare - Bendixson theorem [6] indicates that deterministic
chaos may occur. While numerically integrating Eq. (18), one can
simultaneously calculate the Lyapunov exponents appropriate to the
solution [7].  If the largest exponent is positive, there is
deterministic chaos.

\section{Appendix C}

Consider the quantum mechanical treatment of a macroscopic system.
By definition, such a system is sufficiently large that intensive
properties are invariant to size increase. This suggests
considering such systems in the limit of infinite size and
constant density, know as the \textit{thermodynamic limit}[8],
where phase transitions are sharp (not broadened) and Poincare
recurrences do not occur. In the last three decades there has been
considerable progress in the algebraic treatment of infinite
systems, both with regard to quantum field theory and with regard
to quantum statistical mechanics [8,9].

There can be many representations of the quantum mechanical
operators corresponding to the observables of a physical system.
For a finite system, all representations are \textit{unitarily
equivalent}[10], so that the choice of representation is a matter
of convenience. In contrast, for an \textit{infinite} system there
can be many unitarily \textit{inequivalent} representations.  Each
representation can have associated with it a complete set of
microscopic states and microscopic variables. The various
inequivalent representations can be distinguished by the values of
\textit{intensive macroscopic variables}  associated with them.
For example, in a ferromagnet the components of the magnetization
vector are macroscopic variables. There is a three-fold continuous
infinity of inequivalent representations associated with the
values of these components.

For our purposes, a most important property is that the
Hamiltonian is a functional of the representation [ Ref. 9, p.
32]. A classic example is the BCS theory of superconductivity
[11], where the so-called reduced Hamiltonian contains the complex
superconducting order parameter  $ \Delta $ , whose magnitude is a
measure of the number of Cooper pairs in the system (and thus a
function of temperature) and whose phase plays a crucial role in
the Josephson effect [12].

Since the Hamiltonian is a functional of the representation, and
thus the state of the system, the matrix elements $M_{jk}$ of Eq.
(19) are functions of the coefficients $ a_{i}$  .  Thus the
equations of motion, Eq. (18), are nonlinear in the coefficients,
and deterministic chaos may occur. More specifically, the
Hamiltonian H and the matrix elements $M_{jk}$  are functions of n
macroscopic variables indexing the various inequivalent
representations, the macroscopic variables in turn being functions
of the $ a_{i}$ . Thus the N  dimensional flow of Eq. (18) can be
used to generate an  n dimensional nonlinear flow involving the n
macroscopic variables. The Poincare-Bendixson theorem [6] requires
$n \geq 3$ for the possibility of chaos.

Since the Hamiltonian is a functional of the state of the system,
the superposition theorem breaks down. This may have something to
do with the fact that no one has ever observed a coherent
superposition of the two macroscopic states of Schr\"{o}dinger's
cat [13].

\section{References}

[1]  D. Bohm and B. J. Hiley, \textit{The Undivided Universe}
(Routledge, London 1993).

[2]  P. R. Holland, \textit{The Quantum Theory of Motion}
(Cambridge University Press,Cambridge 1993).

[3]  D. Bohm, Phys. Rev. \textbf{85}, 166 (1952).

[4]  R. H. Parmenter and R. W. Valentine, Phys. Lett. \textbf{ A
227}, 5 (1997).

[5]  R. H. Parmenter and L. Y. Yu, Physica D \textbf{80}, 289
(1995).

[6]  M. W. Hirsch and S. Smale, \textit{Differential Equations,
Dynamical Systems, and Linear Algebra }(Academic Press, New York
1974).

[7]  A. Wolf, J. B. Swift, H. L. Swinney, and J. A. Vistano,
Physica D \textbf{16}, 285 (1985).

[8]  D. Ruelle, \textit{Statistical Mechanics} (W. A. Benjamin,
New York 1969), p. 2.

[9]  G. L. Sewell, \textit{Quantum Theory of Collective Phenomena}
(Oxford University Press,Oxford 1986).

[10]  J. Von Neumann, Math. Annalen \textbf{104}, 570 (1931).

[11]  J. Bardeen, L. N. Cooper, and J. R. Schrieffer, Phys. Rev.
\textbf{108}, 1175 (1957).

[12]  B. D. Josephson, Adv. Phys. \textbf{14}, 410 (1965).

[13]  E. Schr\"{o}dinger in \textit{Quantum Theory and
Measurement}, edited by J. A. Wheeler and W. H. Zurek  (Princeton
University Press, Princeton 1983).

\end{document}